\documentclass{nature}
\usepackage{caption}

\usepackage{enumitem}
\usepackage{graphicx}

\usepackage{nameref}
\usepackage{xr-hyper}
\usepackage{hyperref}
\usepackage{textgreek}
\usepackage{lineno}

\usepackage[]{graphicx,bm,subfigure,amsmath}

\usepackage{placeins}

\makeatletter

\makeatletter

\usepackage[utf8]{inputenc}
\makeatletter
\let\saved@includegraphics\includegraphics
\AtBeginDocument{\let\includegraphics\saved@includegraphics}
\renewenvironment*{figure}{\@float{figure}}{\end@float}
\makeatother

\title{Polaritonic non-locality in ultrastrong light-matter coupling}

\author{Shima Rajabali$^{1,*}$, Erika Cortese$^{2}$, Mattias Beck$^{1}$,  Simone De Liberato$^{2}$, J{\'e}r{\^o}me Faist$^{1}$, Giacomo Scalari$^{1,*}$}

\usepackage[usenames, dvipsnames]{color}
\usepackage{soul}

\begin{document}

\maketitle
\begin{affiliations}
 \item Quantum Optoelectronics Group, Institute of Quantum Electronics, ETH Z{\"u}rich, 8093 Z{\"u}rich, Switzerland
 \item School of Physics and Astronomy, University of Southampton, Southampton, SO17 1BJ, United Kingdom

\end{affiliations}

\begin{abstract}

 \textbf{Sub-wavelength electromagnetic field localization has been central in photonic research in the last decade, allowing to enhance sensing capabilities \cite{Khurgin2017} as well as increasing the coupling between photons and material excitations \cite{Forndiaz2019,FriskKockum2019}.  The ultrastrong light-matter coupling regime in the THz range with split-ring resonators  coupled to magnetoplasmons has been widely investigated, achieving successive world-records for the largest light-matter coupling ever achieved \cite{Bayer2017, Maissen2014, Hagenmueller2010,Scalari2012}. Ever shrinking resonators have allowed to approach the regime of few electrons strong coupling \cite{Keller2017,Ballarini2019}, in which single-dipole properties can be modified by the vacuum field \cite{Cwik2016}.  
 Here we demonstrate, theoretically and experimentally, the existence of a limit to the possibility of arbitrarily increasing electromagnetic confinement in polaritonic systems. Strongly sub-wavelength fields can excite a continuum of high-momenta propagative magnetoplasmons. This leads to peculiar nonlocal polaritonic effects, as certain polaritonic features disappear and the system enters in the regime of bound-to-continuum strong coupling\cite{DeLiberato2017,Cortese2019,Cortese2020}.
 Emerging nonlinearities due to the local breaking of Kohn's theorem are also reported.}

\end{abstract}

Nanophotonic structures confine electromagnetic radiation below the Abbe diffraction limit by storing part of the electromagnetic energy into kinetic energy of moving charges \cite{Khurgin2017}. Primarily relying on metals as charge reservoirs, plasmonics has mainly targeted the visible portion of the electromagnetic spectrum. The possibility to extend the plasmonic excitation to low frequencies (Mid-IR and THz)\cite{StanleyNatPhot2012,Taliercio2019, THZplasmonreview2020} employing semiconductors and two-dimensional systems with tunable plasma frequency through carrier concentration has allowed extreme electromagnetic field confinements in nanostructures \cite{FeiBasovGraphene2012Nature}. 
%Intersubband plasmonic excitations have successfully been coupled to nanometric-sized metallic resonators\cite{Jeannin2020}. 
In parallel, ultrastrong light-matter coupling \cite{Forndiaz2019,FriskKockum2019} has gained recently a lot of attention due to the possibility of observing new quantum phenomena as squeezed vacuum  \cite{Ciuti2005} and long-sought superradiant quantum phase transition \cite{NatafNatComm2009} . Several recent experiments have also allowed to access and measure these squeezed states and their correlation properties\cite{Riek2015,Benea-Chelmus2019}.
Ultrastrong coupling (USC) takes place when the vacuum Rabi frequency ($\Omega_R$), measuring the resonant half-splitting between the lower and upper polariton branches, becomes a considerable fraction of the frequency of the uncoupled systems ($\omega_0$).
Semiconductor platforms allow a high degree of control and flexibility and have proven very successful for the study of this physics\cite{Ciuti2005,Anappara2009,Todorov2010}.  Multiple efforts have been made to scale down the size of photonic resonators \cite{Ballarini2019,Maissen2017}. This scaling aims to increase the light-matter coupling and reach the regime of few electrons strong coupling in molecular \cite{Chikkaraddy2016} and solid-state devices \cite{Reithmaier2004} , where nonlinearities become important and individual electronic degrees of freedom can be manipulated \cite{Todorov2014,Cwik2016}.
 
Landau polaritons \cite{Scalari2012,Paravicini2019} are an experimental platform where cavity photons are (ultra)-strongly coupled to magnetoplasmon excitations occurring in the near field of electronic (LC) metamaterial resonators. They have been demonstrated in electron and hole gases \cite{KellerPRB2020} confined in semiconductor heterostructures. 
The strongest achieved couplings ($\frac{\Omega_R}{\omega_0}>1.4$) make use of metallic split-ring resonators that are able to confine the mm-wave radiation in extremely sub-wavelength volumes \cite{Bayer2017}.  Reducing the capacitor gap in the LC circuit that constitutes the meta-atom has been shown to dramatically enhance the THz fields both for split-ring resonators \cite{Bagiante2015} and non-resonant structures\cite{THZnanoslitSCIREP2014,Bahk2017}. Obtaining the same field enhancement in cavities loaded with active material (i.e., semiconductor quantum wells) requires nevertheless a careful analysis of the different components involved. An important question arises: What are the physical limitations to reducing the cavity volume and subsequent increase of the light-matter coupling?
In this work  we demonstrate how polaritonic nonlocal effects, consisting in the generation of high-momenta magnetoplasmons by the sub-wavelength resonator, effectively limit the achievable field confinement, and thus the resonant polaritonic splitting.
 Our theory shows that bound-to-bound models underlying the polaritonic framework are not valid anymore below a threshold gap size, as the increased momentum uncertainty due to the spatial electromagnetic confinement couples the discrete cavity mode to a continuum of high-momenta magnetoplasmons.
 The system then converts to a bound-to-continuum light-matter coupling model, whose non-perturbative physics has just recently started to be investigated \cite{DeLiberato2017,Forndiaz2017,Cortese2019,Cortese2020}.
 Experiments and finite element simulations support this picture, demonstrating multiple novel nonlocal polaritonic features, predicted by the theory. In particular a reduction of the capacitor gap leads to a progressive disappearance of the upper polariton branch and a vanishing contrast below a threshold magnetic field for the lower branch.

\section*{Polaritonic Nonlocality}
\label{section:theory}
	
Below critical lengthscales, the propagative nature of charge excitations in nanophotonic devices cannot be neglected anymore, leading to the emergence of nonlocal effects, as demonstrated in both plasmonic \cite{Ciraci2012} and phononic \cite{Gubbin2020} systems. In such a regime the nanoscopic features confining the charge distribution, acting as a grating, allow the electromagnetic field to couple with high-momenta matter resonances. These modes can act as loss channels and reduce the field confinement by smearing the distribution of surface charges, thus ultimately limiting the achievable field enhancement.

A different approach is to use the confined electromagnetic field itself to define the coupled region out of an extended electronic system, as in metal nanogap resonators fabricated on the top of a two-dimensional electron gas (2DEG) \cite{Keller2017}. Beyond the comparative ease of designing and tuning electromagnetic nanoresonators, this procedure also tends to maximise the modal overlap between light and matter modes, having contributed to multiple world-records in the achieved coupling strength \cite{Bayer2017,Maissen2014,Scalari2012}. In resonator-defined systems nonlocal effects are also eventually bound to play a role. Tightly bound electromagnetic modes have in fact ill-defined momenta, making the standard momentum-space Hopfield approach inapplicable.  In non-dispersive systems strong light-matter coupling can still be studied using real-space approaches \cite{Gubbin2020}, but when the dispersion becomes non-negligible we are obliged to take into account the 
coupling of the electromagentic field to a continuum of high-momenta propagative electronic modes.  

\begin{figure}
    \centering
    \includegraphics[width=6in]{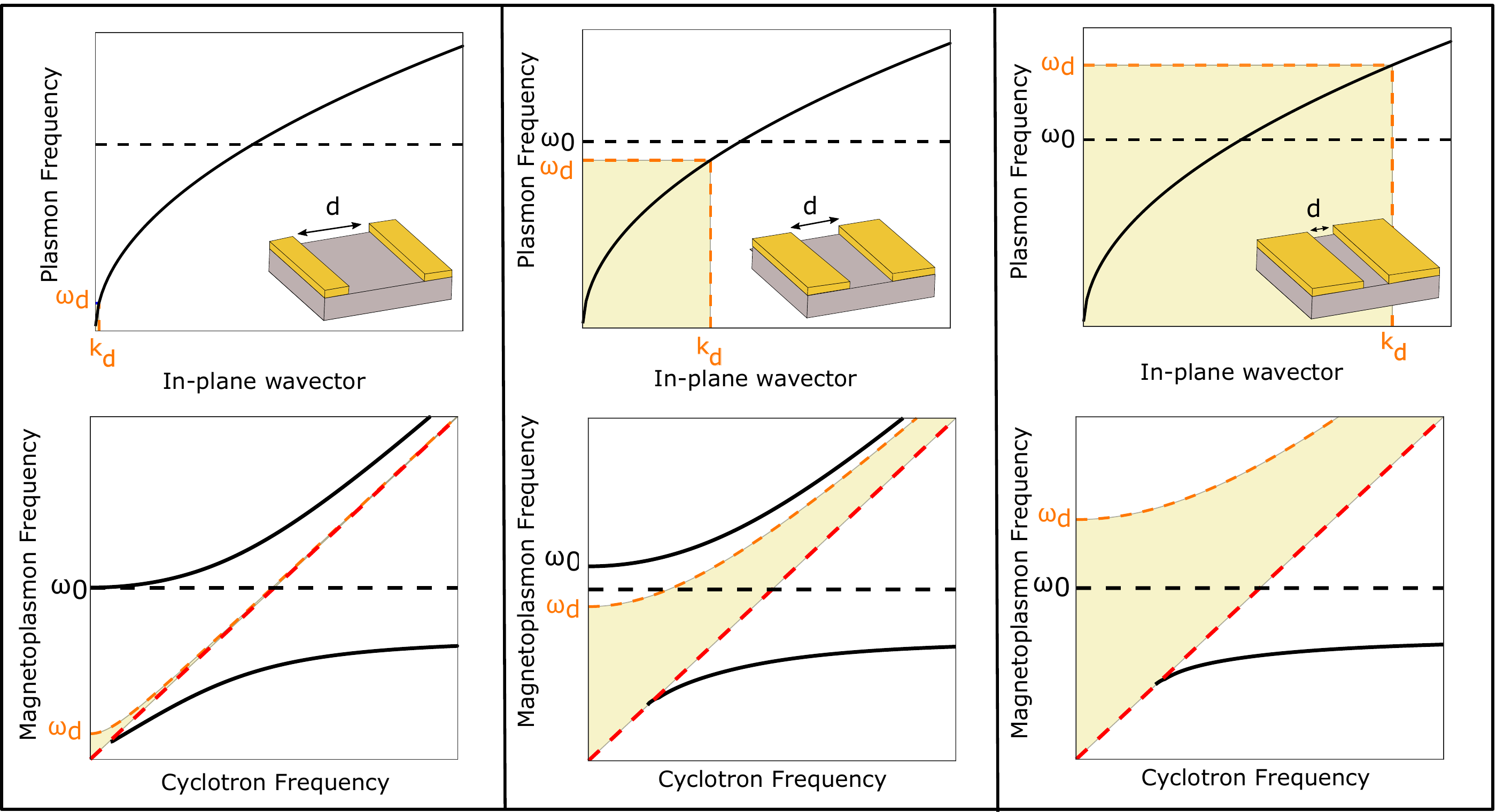}
    \caption{\textbf{Impact of nonlocality in nanoplasmoncs.} 
    Top row: Dispersion of the 2D plasmon. A resonator with features of width $d$ couples with plasmonic waves with in-plane wavevector up to a cutoff $k_d$, corresponding to the frequency range shaded, going from $0$ to a cutoff frequency $\omega_d$.
    Bottom row: Polaritonic resonances as a function of the cyclotron frequency.
    The shaded regions between the cyclotron transition $\omega_c$ (red dashed line) and the magnetically-shifted cutoff  $\bar{\omega}_d=\sqrt{\omega_d^2+\omega_c^2}$ (orange dashed line) highlight the continuum of possible magnetoplasmonic energies to which the photonic resonator (dashed black line) can couple. The insets in the top row show a scheme of the nanogap resonator on the 2DEG, and the three columns are relative to different nanogap sizes, decreasing toward the right.}
    \label{fig:Nonlocality}
\end{figure}

In this section we aim to develop a polaritonic theory able to study nonlocal effects in an extended electronic system
coupled to a photonic nanoresonator, allowing us to understand their impact on the precise engineering of light-matter coupling at the nanoscale. The optical response of a 2DEG in the absence of applied magnetic field can be described in terms of plasma waves indexed by their in-plane momenta $\mathbf{k}$. Their dispersion, shown in the top row of Fig. \ref{fig:Nonlocality}, reads  
\begin{eqnarray} 
\label{eq:omega_P}
		\omega_P^2(k)&=&\frac{k e^2 \rho_{2DEG}}{2m^*\epsilon_0\epsilon_r},
\end{eqnarray}
where $\rho_{2DEG}$ is the 2DEG density, $m^*$ the electron effective mass, and ${\epsilon_r}$ the background effective dielectric function  \cite{Batke1986}. A perpendicular magnetic field $B$, with cyclotron frequency $\omega_c$, will dress the plasmons, leading to field-dependent magnetoplasmon resonances with frequency
$\bar{\omega}_P=\sqrt{\omega_P^2+\omega_c^2}$. Note that in this paper the bar over a frequency will be consistently used to indicate magnetically shifted frequency values.
The modal coupling of the 2DEG to a localised electromagnetic mode with normalized 3D field profile $\mathbf{f(r)}$ and frequency $\omega_0$ can be described by the resonant density of coupling $g(\omega)$, proportional to the 2D Fourier transform of the field profile in the 2DEG plane. 

As shown in Fig. \ref{fig:Nonlocality}, if the smallest geometric feature of the electromagnetic field is of order $d$
the electromagnetic mode will couple to the continuum of magnetoplasma waves with in-plane wavevector up to a cutoff $k_d$ of order $\frac{1}{d}$, covering for $B=0$ a spectrum with cutoff  $\omega_d=\omega_P(k_d)$. Extreme photonic confinement can thus dramatically change the nature of the light-matter system, from a standard Hopfield model describing the strong coupling between discrete light and matter resonances \cite{Hagenmueller2010} (left column of Fig. \ref{fig:Nonlocality}), to the coupling between a discrete photonic mode and a bound-to-continuum matter excitation (right column of Fig. \ref{fig:Nonlocality}), a system in which strong coupling has been only recently achieved \cite{Cortese2020}.
As detailed in the Methods section, the system can then be described by a multimode dissipative bosonic Hamiltonian, similar to the one numerically solved in Ref. \cite{Cortese2019}. 
 
For any fixed value of magnetic field, such a model describes a photonic resonance coupled to a continuum of magnetoplasma waves, starting at the cyclotron frequency $\omega_c=\bar{\omega}_P(k=0)$ and vanishing for frequencies substantially larger than $\bar{\omega}_d=\bar{\omega}_P(k_d)$. The continuum part of the spectrum is shaded in Fig. \ref{fig:Nonlocality}.
The two Landau polariton branches \cite{Scalari2012}, shown by black solid lines, are strongly modified when $\omega_d$ becomes non-negligible compared to other frequencies of the problem. While the two polaritonic branches normally exist for any value of the magnetic field in a standard polaritonic model \cite{Hagenmueller2010}, our theory predicts these modes to disappear for finite values of the parameters, where the narrow polaritonic resonance vanishes into the continuum region. 

We find that the region in which the lower polariton exists satisfies the equation
\begin{eqnarray}
\label{eq:ex_LP}
		\int_0^{\infty}d\omega
		\frac{\lvert g(\omega)\rvert^2}{{\omega}^2}>\frac{{\omega}_0^2-\omega_c^2}{{4\omega_c^2}}.
\end{eqnarray}
 The left hand side of Eq. \ref{eq:ex_LP} is always positive and the condition is thus always satisfied for $\omega_c>\omega_0$. In this case the lower polariton, which starts at $\omega_0$ when the light-matter coupling is neglected, red-shifts for larger couplings and it never crosses into the continuum region starting at $\omega_c$. The same equation is instead never verified for $\omega_c=0$. The lower polariton branch is thus interrupted at a finite value of the cyclotron frequency. Assuming that the coupling $g(\omega)$ vanishes for  frequencies larger than $\omega_d$, we also proved that the upper polariton exists if and only if $\omega_0>\omega_d$, independent of the value of the magnetic field. The latter inequality can be very simply understood as diffraction condition requiring that the wavelength of the plasma waves must be larger than the gap width. This is consistent with the results shown in Fig. \ref{fig:Nonlocality}, where the two polaritonic resonances are present in the left column, while in the central one the upper polariton is still present but the lower polariton disappears for a finite value of the cyclotron frequency. For even smaller nanoresonators also the upper polariton completely disappears, as shown in the right column.

\section*{Experimental Results}
\label{section:experiment}

\begin{figure}
   \centering
    \includegraphics[width=4.2in]{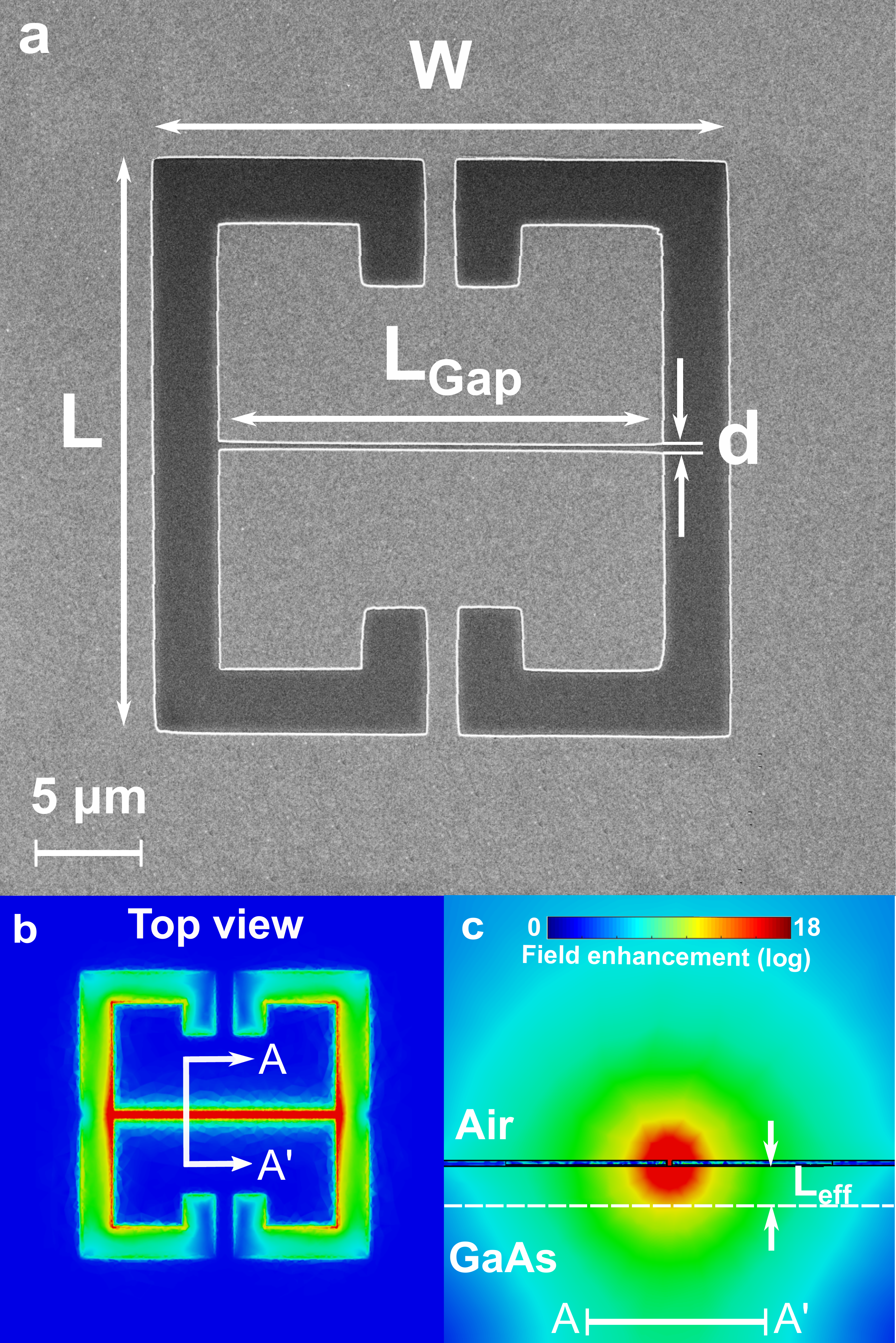}
   \caption{\textbf{Cavity and electric field parameters} (a) A SEM image of the cSRR with 250 nm gap showing different resonator parameters including gap size ($d$). (b,c)  Cold cavity electric field distribution, simulated by CST, for cSRRs resonating at 500 GHz in two different views: Top view (part b) and AA' view or yz plane (part c). You can find $L_{eff}$ in part c. Part b and c have the same colorbar.}
   \label{fig:sem}
   \end{figure}

\begin{figure}
    \centering
    \includegraphics[width=6in]{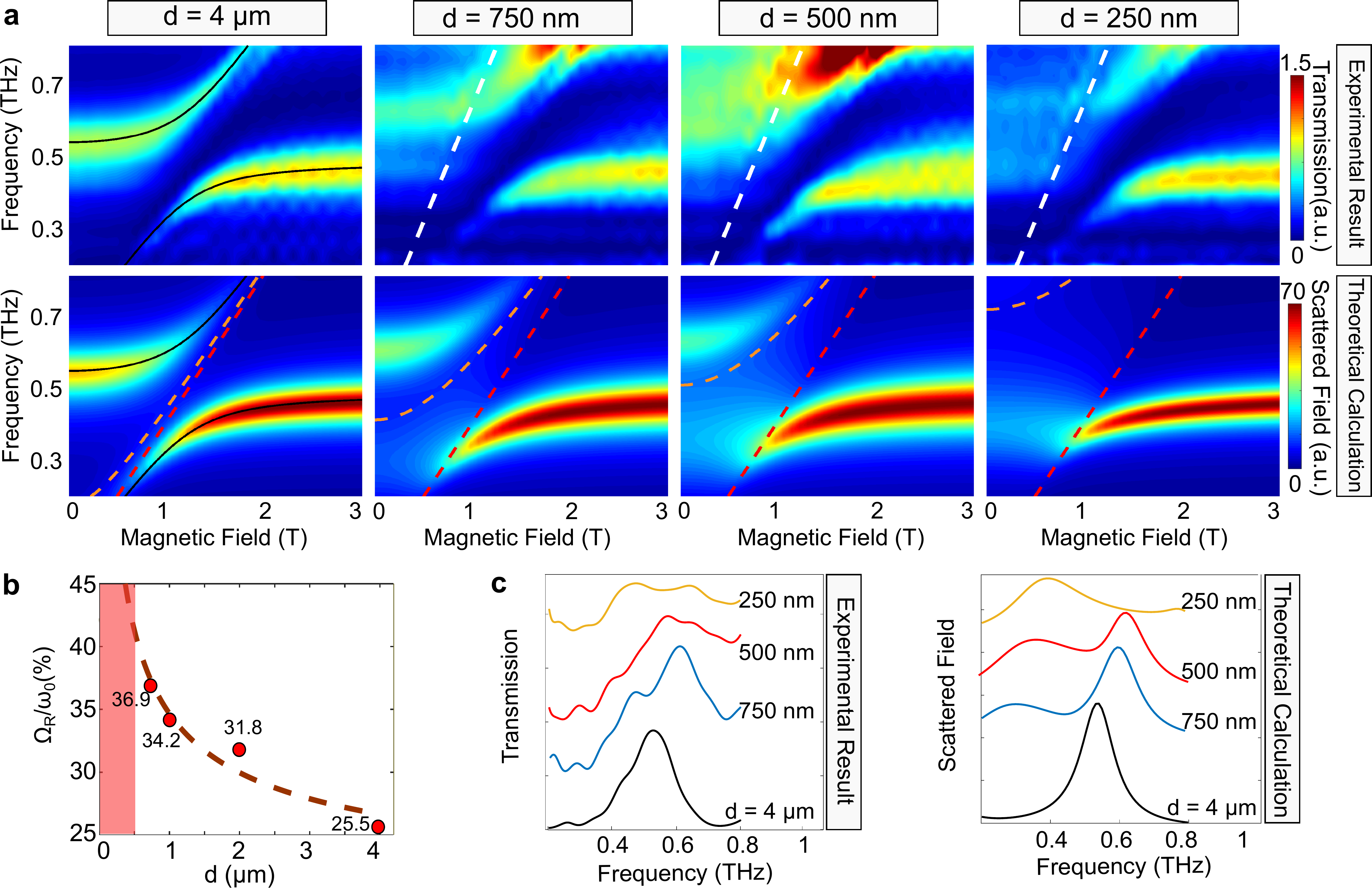}
    \caption{\textbf{Theoretical vs. Experimental results} (a) Top row: transmission of cSRRs coupled to Landau level transition in 2DEG vs. magnetic field for four different gap sizes (d). Bottom row: the calculated scattered Field for a gap feature size of $d$, same as gap sizes in experiment result. The black solid lines are fitted LP and UP. The dashed white line shows an additional feature corresponds to $f = 1.6 \times \omega_c /2\pi$. The regions between the cyclotron transition $\omega_c$ (red dashed line) and the magnetically-shifted cutoff (orange dashed line) are similar to shaded regions on the bottom row of  Fig. 1, showing the region of continuum magnetoplasmon energies (b) Normalized coupling vs. $d$. Dashed brown line indicates the predicted trend of $\frac{1}{\sqrt{d}}$. The red region spans the gap where the broadening of the upper branch makes the measurement of the coupling meaningless.(c) Sections (offset for clarity) of the color plots at zero magnetic field for experimental result and theoretical calculation in part a. The progressive upper branch broadening is evident.}
    \label{fig:measurements_gapsweep}
\end{figure}

We now proceed to the experimental study where we measure a 2DEG coupled to complementary split-ring resonators (cSRR)s for different values of the central gap width $d$, from 4 \textmu m down to 250 nm. The sample transmission is measured using THz time-domain spectroscopy (TDS) at temperature T = 2.7 K  as a function of the magnetic field swept between 0 and 4 T. %More detail about the measurement setup is available in Ref. \cite{Scalari2012}.
A scanning electron microscope (SEM) picture  of the resonator  is shown in Fig.~\ref{fig:sem}a. \\
In cold cavity case, the cSRRs' fundamental mode electric field is concentrated in the central gap and its modal volume can thus be written as $V = L_{Gap}\times d \times L_{eff}$, where $L_{eff}$ is the effective penetration depth of the fringing field inside the substrate (Fig.~\ref{fig:sem}c). 
Note that, as the central gap acts as a capacitor in the LC split-ring circuit, it directly affects the resonant frequency of the resonator $\omega_0 =2\pi f_{LC}= \frac{2\pi}{\sqrt{LC}}$. In order to allow for a meaningful comparison between different samples we thus rescaled the resonators in order to keep a fixed cSRR frequency $f_{LC}=500$GHz.
Only the electrons below the cavity effective surface $S = L_{Gap}\times d $ are coupled to the resonator, leading to a total effective electron number $N_{2DEG}=\rho_{2DEG}S$. 
The vacuum Rabi frequency $\Omega_R\propto\sqrt{\frac{N_{2DEG}}{V}}$, quantifying resonant coupling of a discrete photonic mode to a single magnetoplasmon resonance, can then be explicitly written as\cite{Maissen2014} 
\begin{equation}
  \Omega_R =  \sqrt{\frac{e^2 \rho_{2DEG}}{4m^*\epsilon_0 \epsilon_r  L_{eff} }}.\label{eq:1}
\end{equation} 
Although the explicit dependence of the coupling upon the gap width $d$ cancels in Eq.~\ref{eq:1}, $L_{eff}$ changes due to the reduced dimension. This dependency is borne out by our finite element electromagnetic simulations for the cold cavity case, where there is no 2DEG underneath the metallic surface, showing a more confined and enhanced electric field in a narrower gap (supplementary info). Reducing the gap size $d$ is thus expected to increase the coupling $\Omega_R$ for a fixed cavity frequency \cite{Maissen2014}.
This is verified by Fig.~\ref{fig:measurements_gapsweep}b , where we plot the the normalized coupling ratio $\frac{\Omega_{R}}{\omega_0}$  extracted from a theoretical fit of the data for $d>750$nm, where the nonlocal effects discussed above are not relevant and the system is well described by a standard single mode Hopfield model\cite{Hagenmueller2010}. In such a regime we can extract a $\frac{1}{\sqrt{d}}$ dependence of the normalized coupling over the gap width (dashed brown line).

\newpage

\par In the top row of Fig.~\ref{fig:measurements_gapsweep}a we report the experimental results. The colormap corresponding to $d=4$  \textmu m gap size ($\omega_0\gg \omega_d$) shows an anti-crossing of the first cavity mode with the linear cyclotron transition dispersion at an out plane magnetic field B $\approx$ 1.2 T. The solid black lines show the lower polariton and upper polariton branches fitted to the extracted maximum of transmission using our theory in the bound-to-bound regime, equivalent to the standard Hopfield model \cite{Hagenmueller2010, Scalari2012}.

For the resonator with d=750 nm ($\omega_0>\omega_d$) we observe a broadening of the UP branch but we can still extract the maximum of the transmission at each magnetic field. For 500 nm ($\omega_0\approx \omega_d$) and especially for 250 nm ($\omega_0< \omega_d$) gap sizes the upper branch does not display a clear maximum anymore and the signal is broadened over a frequency range of over 200 GHz. \\
Our optical measurements probe the photonic part of the polaritonic excitations, which for the lower branch naturally vanishes when $B\rightarrow 0$. This effect makes it difficult to highlight the lower polariton disappearance at finite values of the magnetic field predicted by our theory. Inspection of \ref{fig:measurements_gapsweep}a nevertheless shown a clear reduction of the lower branch visibility range for the smallest gap.

In order to better compare to experimental results, Fig. \ref{fig:measurements_gapsweep}a (bottom row) plots theoretical calculation results obtained including losses affecting both the photonic and magnetoplasmon fields.
As detailed in the Methods section, this has been accomplished extending the dissipative diagonalization procedure for cavity quantum electrodynamics (CQED) from Ref. \cite{DeLiberato2017}. The cyclotron transition $\omega_c$ (red dashed line) and the magnetically-shifted cutoff $\bar{\omega}_d=\sqrt{\omega_d^2+\omega_c^2}$ (orange  dashed  line) bound the region of continuum magnetoplasmon energies 
 shaded  in Fig. \ref{fig:Nonlocality} and they visualize the loss channel leading the broadening of the upper polariton branch in our transmission data.
\\
The enhanced transmission visible at higher frequencies in the experimental data is due to the second mode of the cSRRs lying close to $1$THz. This mode is not strongly confined in the gap and it is not relevant for the nonlocal physics of the system. It has thus not been included in the theoretical modeling.\\
To clearly illustrate the upper polariton broadening, sections of all four color plots for both experimental and theoretical results in Fig.~\ref{fig:measurements_gapsweep}a at zero magnetic field are plotted (Fig.~\ref{fig:measurements_gapsweep}c). The progressive broadening of the upper polariton branch is manifest.
The clear blue shift of the upper polariton from 4 \textmu m to 750 nm gap at zero magnetic field is related to the opening of the polariton gap \cite{Todorov2010}, another indication of larger coupling strength in cSRRs with small slits \cite{Maissen2014}.

Other important features in the  experimental transmission measurements  are the linear dispersions appearing for the smallest gaps in Fig. \ref{fig:measurements_gapsweep}a. These features can be fitted by a linear function f = 1.6 $\times \omega_c/2\pi$ (white dashed lines) that corresponds to a second cyclotron harmonic of a 2DEG with an effective mass 8 percent heavier than the one used in this study. A heavier effective mass can appear in the systems where the translational invariance breaks and Kohn's theorem\cite{Kohn1961} does not hold anymore. The translational invariance may not hold when the gap size reduces down to few hundreds of nm, as already anticipated theoretically and observed for the grating-coupled system of Ref. \cite{Batke1986}. These features are not reproduced of course neither by the classical modeling used in the finite element calculations, nor by our linear quantum theory. The presence of harmonics of the cyclotron transition in metamaterial-coupled Landau polaritons was already observed by some of us in 2DHG \cite{KellerThesis2018}).

\section*{3D finite element electromagnetic simulations}
In order to quantitatively confirm the theoretical predictions using the detailed cSRR geometry and visualize the nonlocal coupling to 2D magnetoplasmon modes, we recurred to finite element electromagnetic simulation (CST Microwave Studio) in which the 2DEG is modeled as a gyrotropic material. No ``effective medium'' is adopted, i.e., the dimensions of the different components including the thickness of the quantum well layer (10 nm) are kept as in reality. The electron density of the simulated 2DEG is calibrated through a simulation of the 2DEG only and compared with the real measurements performed on the heterostructure used in the experiment.
\par
\begin{figure}
    \centering
    \includegraphics[width=6in]{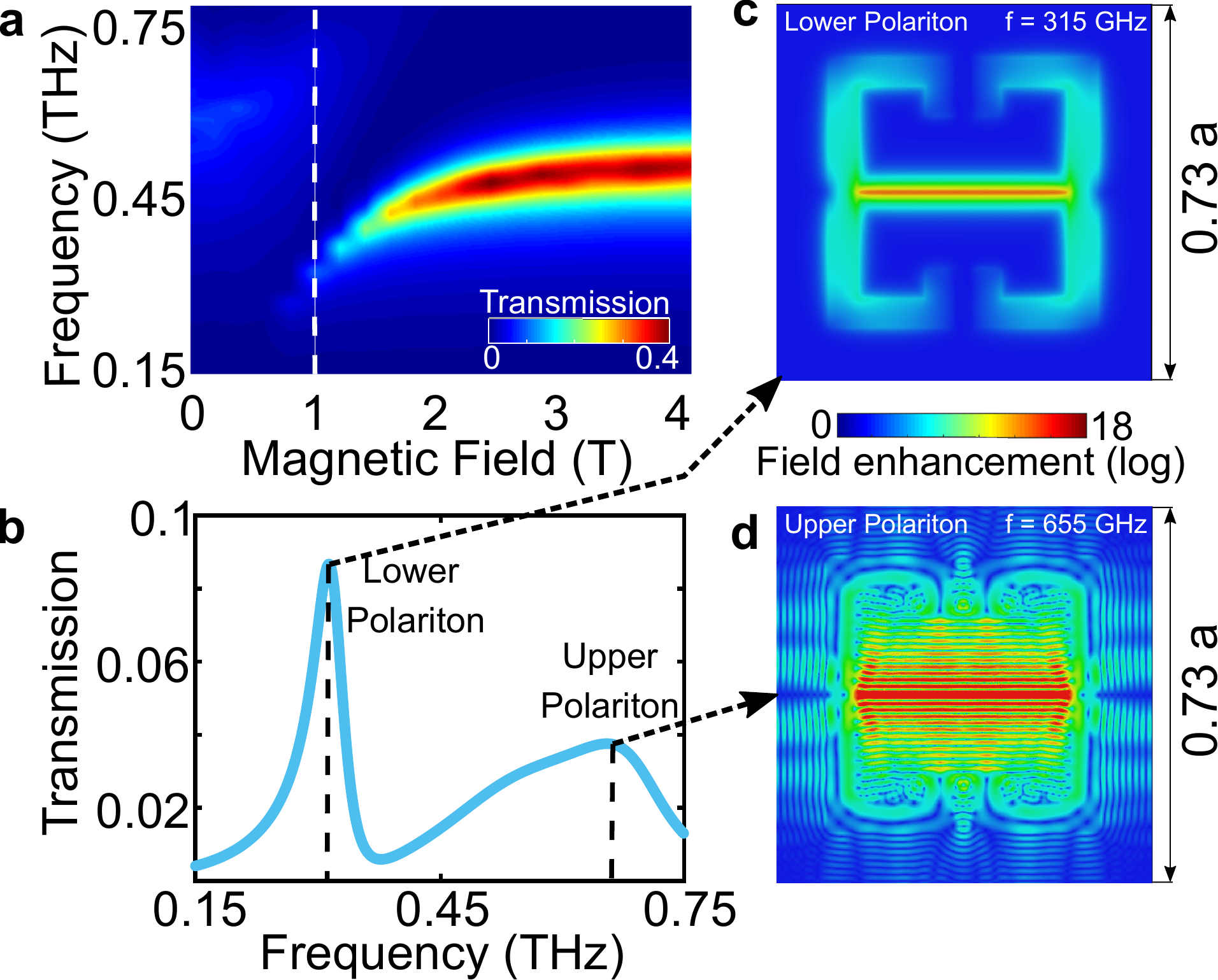}
    \caption{\textbf{Finite element simulation for cSRR with 250nm gap on 2DEG} (a) Simulation of the Landau polariton dispersion as a function of magnetic field. The broadening and loss of contrast of the upper branch is evident (b) section of the color plot of panel (a) at the anti-crossing field B=1 T. In panels (c) and (d) we report the electric field distributions for  UP and LP. The field distribution in (c) for LP shows a pronounced electric field concentration in the 250 nm gap and a corresponding narrow spectral distribution. On the other side part (d) clearly demonstrates the excitation of plasmonic standing waves and the corresponding peak in the frequency spectrum is broad.}
    \label{fig:upperbranchnrrowgap}
\end{figure}

 A set of simulation of the Landau polariton dispersion for the metallic metasurface with a gap of 250 nm deposited on 2DEG as a function of magnetic field is reported in Fig.\ref{fig:upperbranchnrrowgap}a. In this simulation, magnetic field is swept in the range B $\in [0,4] $ T,  and can be directly compared with the corresponding experimental plot of Fig. \ref{fig:measurements_gapsweep}a. As visible, the finite element simulation reproduces very well the broadening of the upper polariton branch. 
It is interesting to inspect the electric field distribution at the anti-crossing (white dashed line in Fig. \ref{fig:upperbranchnrrowgap}a). As visible from part \ref{fig:upperbranchnrrowgap}d the upper branch, expected to sit in the magnetoplasmon continuum, is completely dominated by the propagative plasmonic behavior and the corresponding peak is largely broadened. Conversely, the lower branch (part \ref{fig:upperbranchnrrowgap}c) displays the electric field intensity all concentrated in the gap.  More simulation data can be find in our supplementary info.

\section*{Conclusions}
In conclusion, we theoretically investigated the emergence of nonlocality in polaritonics. Contrary to nonlocality in plasmonic \cite{Ciraci2012} and phononic \cite{Gubbin2020} systems, caused by tight charge confinement, here nonlocal effects are driven by the confinement of the resonator's electromagnetic field coupling to a continuum of propagating magnetoplasma excitations. We experimentally observed such a physics using Landau polaritons at sub-THz frequencies employing nanometer-sized resonators. 
First, we show that when nonlocal effects are not relevant (gap size $d>750$ nm) the normalized coupling ratio between light and matter scales as $\frac{1}{\sqrt{d}}$, increasing by about 50$\%$ from d= 4 $\mu$m to 750 nm, and reaching 37$\%$ for a single QW due to the strong field enhancement. 
For smaller gap values, nonlocal effects become dominant and the system is not anymore well described by a standard Hopfield model. As predicted by our multimode dissipative bosonic Hamiltonian we observe a broadening of the upper polariton branch and the partial disappearance on the lower one. Finite element electromagnetic simulations confirm this interpretation. The presence of highly localized electric fields is as well testified by emerging non-linearities breaking Kohn's theorem.
Our findings set quantitative limits to the miniaturization of polaritonic devices, and to the enhancement of polariton gaps, as well as open new possibilities in the study of bound-to-continuum strongly coupled systems \cite{DeLiberato2017,Cortese2019,Cortese2020}.

\section*{Methods}\label{method}

\subsection{Theoretical description of light-matter coupling.}
\label{theory}

We model the magnetoplasmonic excitations coupled to a single photonic mode using a generalisation of the multimode Hamiltonian developed in Ref. \cite{Cortese2019} 
\begin{eqnarray} \label{eq:Hd}
		H&=&\hbar\tilde{\omega}_0 a^{\dagger}a+\sum_{\mathbf{k}}\,\hbar\bar{\omega}_P(k) \,b_{\mathbf{k}}^{\dagger}b_{\mathbf{k}}
		+i {\hbar\Omega_R}\sum_{\mathbf{k}} \sqrt{\frac{\bar{\omega}_P(k)}{\tilde{\omega}_0}} \left[a^{\dagger}+a \right]
		\left[\Xi_{\mathbf{k}} b_{\mathbf{k}}^{\dagger}-\Xi_{\mathbf{k}}^*b_{\mathbf{k}}\right],
\end{eqnarray} 
where the $a$ and $b_{\mathbf{k}}$ are annihilation operators for a resonator photon and for a magnetoplasma wave with in-plane wavevector $\mathbf{k}=(k_x,k_y)$ and bare ($\omega_c=0$) frequency $\omega_{k}$. The second quantised operators obey bosonic commutation relations $\left[ a,a^{\dagger}\right]=1$ and $\left[ b_{\mathbf{k}},b_{\mathbf{k'}}^{\dagger}\right]=\delta_{\mathbf{k,k}'}$, where the $\delta$ is a Kronecker symbol. As better detailed in the supplementary info, the photonic frequency $\tilde{\omega}_0$ includes the renormalisation due to the diamagnetic $A^2$ term in the Hamiltonian, while $\bar{\omega}$ is the magnetoplasmon frequency. 
In Eq. \ref{eq:Hd} the vacuum Rabi Frequency $\Omega_R$ is given by Eq. \ref{eq:1} and $\Xi_{\mathbf{k}}$ is the 2D Fourier transform of the resonator field in the 2DEG plane.
In the continuum regime the Hamiltonian in Eq. \ref{eq:Hd} can be rewritten
\begin{eqnarray} \label{eq:H}
		H&=&\hbar\tilde{\omega}_0 a^{\dagger}a+\int_0^{\infty} d\omega\,\hbar\bar{\omega} \,b(\omega)^{\dagger}b(\omega)
		+i \hbar\int_0^{\infty}d\omega \sqrt{\frac{\bar{\omega}}{\tilde{\omega}_0}} \left[a^{\dagger}+a \right]
		\left[g(\omega)b(\omega)^{\dagger}-g(\omega)^*b(\omega)\right],
\end{eqnarray}
with $\left[ b(\omega),b(\omega')^{\dagger}\right]=\delta(\omega-\omega')$, and the $\delta$ is now the Dirac function.
	The secular equation providing the discrete eigenvalues of the Hamiltonian in Eq. \ref{eq:H} can be written as
	\begin{eqnarray}
\label{eq:up_ex}
		\int_0^{\infty}d\omega'
		\frac{\lvert g(\omega')\rvert^2}{{\omega}^2-\bar{\omega}'^2}=\frac{{\omega^2-\omega}_0^2}{4\omega^2}.
\end{eqnarray}
	By evaluating such an equation on the boundaries of the continuum region ($\omega=\omega_c$ and $\omega=\bar{\omega}_d$) it is then possible to derive extrenal conditions for the existence of the upper and lower polaritons.
Numerical results are obtained assuming that the only relevant scattering feature, with a cutoff wavevector $k_d=\frac{1}{d}$, is across the $x$ axis. We can then write $\Xi_{\mathbf{k}}=\Theta(k_d-|k_x|)\delta_{k_y,0}$, with $\Theta$ the Heaviside function, leading to a resonant coupling density of the form
	 \begin{eqnarray}
		g(\omega)&=&\Omega_R \sqrt{\frac{2\omega}{\pi \omega_d^2}}\Theta(\omega_d-\omega).
	\end{eqnarray}

\subsection{Theoretical description of the open system}
\label{theory2}
In order to better predict experimental features due to nonlocal nanopolaritonic effects, we need to consider the impact of intrinsic polariton losses, due to the finite lifetime of both the cavity photon and the magnetoplasmon. We used to this aim the approach initially introduced by Huttner and Barnett to quantise the electromagnetic field in bulk dissipative dielectrics \cite{Huttner1992} and recently extended to the CQED case with both material and photonic losses  \cite{DeLiberato2017}.
This approach allows us to to write the system Hamiltonian in terms of a continuum of broadened 
photonic and magnetoplasmonic modes, with annihilation operators $A(\omega)$ and $B(\omega)$,
\begin{eqnarray}\label{eq:Htot}
H_{\textrm{tot}}&=&  \int_0^\infty \hbar\omega A^\dagger(\omega)A(\omega) + \int_0^\infty \hbar\omega B^\dagger(\omega) B(\omega) \\
&&+\hbar \int_0^\infty d\omega  \int_0^\infty d\omega'  \left [ \zeta(\omega) A^\dagger(\omega) + \zeta^*(\omega) A(\omega)  \right] \times \left [ \theta(\omega') B^\dagger(\omega')+ \theta^*(\omega') B(\omega') \right],\nonumber
\end{eqnarray}
satisfying bosonic commutation relations $\left[A(\omega),A(\omega')^{\dagger} \right]=\left[B(\omega),B(\omega')^{\dagger} \right]=\delta(\omega-\omega')$.
In the Hamiltonian above the function $\zeta(\omega)$ contains the information about the photonic losses, and  $\theta(\omega)$ contains a convolution integral between the resonant coupling density $g(\omega)$ and a function describing the frequency-dependent matter losses.

The Hamiltonian in Eq. \ref{eq:Htot} is then  diagonalised by introducing operators for two orthogonal polariton branches $j=\pm$, obeying Bosonic commutation relations
$\left[P_j(\omega),P_{j'}(\omega')\right]=\delta(\omega-\omega') \delta_{j,j'}$,
which we write in terms of the bare broadened fields
\begin{eqnarray} \label{P}
P_{j}(\omega)=\int_0^\infty d\omega' \left [{x}_j(\omega,\omega') A(\omega')+ {z}_j(\omega,\omega') A^\dagger(\omega') + {y}_j(\omega,\omega') B(\omega') +{w}_j(\omega,\omega') B^\dagger(\omega') \right]. 
\end{eqnarray}
The frequency dependent Hopfield coefficients $(x_j,z_j,y_j,w_j)$ can then be found by solving the equations of motion  $\omega P_j(\omega)=\frac{1}{\hbar}\left[ P_j(\omega), H_{\textrm{tot}} \right ]$.

As shown in Ref. \cite{DeLiberato2017} such a description is defined modulo a real function of $\omega$, fixing the basis used to specify the two degenerate modes $P_{\pm}(\omega)$. The two polaritonic operators are thus not individually identifiable with the two polaritonic branches, and the observable results have to be calculated from the gauge-invariant bare electromagnetic mode 
 \begin{eqnarray}
(a+ a^\dagger)&=&  \int^{\infty}_0  d\omega  \sum_{j=\pm} M_j(\omega) \left[ P_j(\omega)+ P_j^\dagger(\omega) \right],
\end{eqnarray}
with $M_j$ the electric component of the polaritonic field obtained inverting the Hopfield transformation. 

\subsection{Sample fabrication, measurement setup, and simulated structure in CST}\label{num1} This part is available in Supplementary info.

\begin{addendum}

\item[Data Availability Statement] The numerical simulation and measurement data that support the plots within this paper are available from the corresponding author upon reasonable request.

\item[Code availability statement] The codes used in the theory part of this study are available from the corresponding author upon reasonable request.
\end{addendum}

\bibliography{Ref.bib}

\begin{addendum}

 \item[Acknowledgements] G.S. would like to thank M. Jeannin and R. Colombelli for discussions. S.R. thanks I.-C. Benea-Chelmus for fruitful discussions. 
 G.S. and  J.F. acknowledge financial
support from the ERC Advanced grant Quantum Metamaterials
in the Ultra Strong Coupling Regime (MUSiC) with the ERC
Grant 340975. G.S. and J.F. also acknowledge financial support
from the Swiss National Science Foundation (SNF) through
the National Centre of Competence in Research Quantum
Science and Technology (NCCR QSIT).
S.D.L. is a Royal Society Research Fellow and was partly funded by the Philip Leverhulme Prize of the Leverhulme Trust. E.C. acknowledges funding from the RPG-2019-174 grant of the Leverhulme Trust.
 \item[Competing Interests] The authors declare that they have no competing financial interests.
 \item[Authors contributions] G.S., J.F. and S.D.L. conceived the idea.  S.R. designed and fabricated the devices, and carried out all the optical measurements, analysed all experimental data and performed numerical simulations under the supervision of G.S. and J.F.. E.C. and S.D.L. developed the theory. M.B. grew the GaAs/AlGaAs quantum wells. S.R., S.D.L. and G.S. wrote the manuscript. All authors discussed the results and commented on the manuscript.
 
 \item[Correspondence]  *Correspondence should be addressed to S. Rajabali (email: shimar@phys.ethz.ch) and G. Scalari (email: scalari@phys.ethz.ch).
\end{addendum}

\end{document}